\documentclass[prb,showpacs,amsmath,amssymb,twocolumn]{revtex4}
\usepackage{graphicx}
\usepackage{amsmath}
\usepackage{amssymb}

\begin{document}

\title{Single-Electron Transistor in Strained Si/SiGe Heterostructures}

\author{Thomas Berer, Dietmar Pachinger, Georg Pillwein, Michael M\"uhlberger, Herbert Lichtenberger, 
Gerhard Brunthaler and F. Sch\"affler}

\affiliation{Institute of Semiconductor and Solid State Physics, Johannes Kepler University Linz, Altenbergerstra\ss{}e 69, 4040 Linz, Austria}

\begin{abstract}
A split gate technique is used to form a lateral quantum dot in a two-dimensional electron gas of a modulation-doped silicon/silicon-germanium heterostructure. e-beam lithography was employed to produce split gates. By applying negative voltages to these gates the underlying electron gas is depleted and a lateral quantum dot is formed, the size of which can be adjusted by the gate voltage. We observe single-electron operation with Coulomb blockade behavior below 1K. Gate leakage currents are well controlled, indicating that the recently encountered problems with Schottky gates for this type of application are not an inherent limitation of modulation-doped Si/SiGe heterostructures, as had been speculated.
\end{abstract}


\maketitle

With the introduction of the Si/SiGe heterobipolar transistor into large scale production, Si-based heterostructures have become an important material system for electronic high-performance device with full compatibility with standard Si technologies \cite{schaeffler}. Meanwhile, the first digital CMOS circuits with selectively grown SiGe epilayers are commercially available, \cite{thompson} and intense research and development is dedicated to the fabrication of SiGe pseudosubstrates and SiGe-on-insulator substrates for further device applications. In terms of basic research, great effort is directed toward spintronic and quantum computing as potential techniques for future computation and encrypting facilities \cite{kane,vrijen}. Si-based heterostructures have distinct advantages in these fields because of the extremely long spin coherence times \cite{wilamowski,lyon}, which are attributed to the small spin-orbit interaction, and the low natural abundance of isotopes with nuclear spin. Moreover, in this material system the interaction between the spins of the conduction electrons and the nuclei can be further reduced or even completely  suppressed by employing enriched $^{28}$Si, and Ge with a depleted $^{73}$Ge isotope. This way, matrix materials free of nuclear spins are conceivable in addition to the unrivalled ultra-large scale integration capabilities of Si technologies. \\

On a device level, single electron transistors (SET) with carrier confinement in all three directions of space are considered a key-component for spin manipulation and programmable entanglement \cite{friesen}. While SET development for laboratory applications have reached a mature state in heterosystems based on III-V compound semiconductors, only few Si/SiGe SETs have been reported \cite{notargiacomo,klein,yablonovitch}. Moreover, none of these was achieved by the usual split-gate technique that is considered a precondition for efficient coupling of quantum dots and for high integration. It was pointed out in some of these papers that a lateral quantum dot cannot be achieved by Schottky split gates due to detrimental leakage currents through threading dislocations in the crystal and Fermi level pinning. Here we demonstrate that excessive leakage currents are not an intrinsic limitation of modulation-doped Si/SiGe heterostructures: Our split-gate SETs show well-behaved Coulomb blockade and very low leakage currents of the Pd Schottky gates.\\

A high-mobility n-type modulation-doped Si/SiGe heterostructure, similar to the one described in detail in Ref. \cite{tsui}, was grown by molecular beam epitaxy (MBE) in a Riber SIVE 45 Si aperatus. In brief, on a standard 4'' Si (001) substrate a 2.5\,$\mu m$ thick relaxed step graded buffer (Si$_{0.95}$Ge$_{0.05}$ to Si$_{0.75}$Ge$_{0.25}$) is grown. A 0.5\,$\mu m$ thick constant
composition buffer (Si$_{0.75}$Ge$_{0.25}$) follows the step graded buffer. The 2DEG is formed at the upper interface of a 150\,\AA{}
thick strained Si channel which is grown subsequently. The 200\,\AA{} thick antimony-doped Si$_{0.75}$Ge$_{0.25}$ layer was grown at 300\,$^{\circ}$C, and is separated from the channel by a 150\,\AA{} thick Si$_{0.75}$Ge$_{0.25}$ spacer layer. Finally a 450\,\AA{} thick Si$_{0.75}$Ge$_{0.25}$ layer and a 100\,\AA{} Si cap were grown at an temperature of 600\,$^{\circ}$C.
Shubnikov-de-Haas and Hall measurements performed at 1.5\,K show an electron density of
$3.2\times{}10^{11}\,cm^{-2}$ with an electron mobility of 150000\,$cm^2/Vs$.

\begin{figure}
\begin{center}\leavevmode
\includegraphics[width=0.7\linewidth]{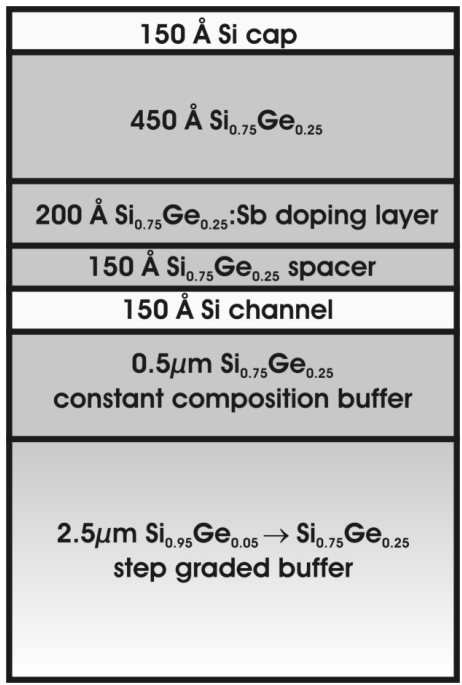}
\caption{Sample Structure}
\label{psg1417}
\end{center}\end{figure}

\begin{figure}
\begin{center}\leavevmode
\includegraphics[width=0.7\linewidth]{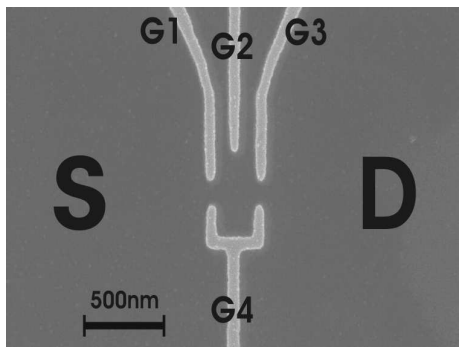}
\caption{Scanning electron micrograph of the palladium split gates. The pitch between the upper gates is 185nm}
\label{set}
\end{center}\end{figure}

Ohmic contacts were formed by deposition of Au/Sb and subsequent annealing at $350\,^{\circ}C$ for 60sec. A Hall-bar structure was prepared by optical lithography and reactive ion etching (RIE) with SF$_6$. Subsequently the split gate structures were written by e-beam lithography with a LEO Supra 35 scanning electron microscope (SEM) into PMMA. The split gates were fabricated by using a lift-off technique. Pd was used as a gate metal, which has (together with Pt) the highest Schottky barrier on n-silicon. A SEM micrograph of the final structure is shown in figure \ref{set}. The pitch between the upper gates is 185\,nm. Finally, connections from the split gates to the bond pads where also made of Pd using optical lithography and lift-off in acetone.\\

As pointed out above, recent papers speculate that a lateral quantum dot cannot be made by the split-gate techniques due to threading dislocations from the relaxed SiGe buffer, which may cause too high leakage currents. To measure the leakage current all gates were connected in parallel to increase the area of the Schottky gates and thus probe the worst-case  leakage currents. Measurements where performed in an He$^3$ cryostat at a temperature of about 300$\,$mK. Down to a voltage of about -3$\,$V leakage currents where below 0.02$\,$nA, which is at the sensitivity limit of the experimental set-up. Below -3$\,$V the current increases rapidly as expected from a Schottky diode. However, conductance oscillations are observed at gate voltages between -1.46 and -1.6\,V, which is in the non-conduction part of the diode characteristics.

\begin{figure}
\begin{center}\leavevmode
\includegraphics[width=0.7\linewidth]{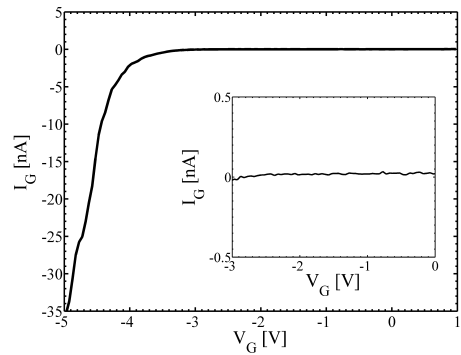}
\caption{For the measurement of the gate IV-characteristics all gates were connected in parallel to
    enhance possible leakage currents. Down to about -3$\,$V the leaking currents are below the measurement
    accuracy. Below -3V currents increase as expected for a Schottky diode. The insert shows a zoom-in of the linear
    range.}
\label{IVcurve}
\end{center}\end{figure}

\begin{figure}
\begin{center}\leavevmode
\includegraphics[width=0.7\linewidth]{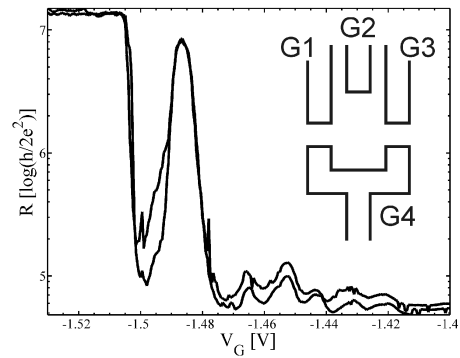}
\caption{Resistance oscillations measured at 300\,mK by changing gate voltages of gates G1 and G2 with fixed gate voltage at G3 and G4. One can easily distinguish different resistance minima separated by resistance peaks in the Coulomb-blockade regions.}
\label{Coul_osz}
\end{center}\end{figure}

By applying negative voltages to the gates the underlying 2DEG can be depleted and a laterally constricted area of free carriers ("dot") is formed between the gates. By varying the plunger gate (G2) voltage, the energy levels inside the
quantum dot can be moved into and out of resonance with the Fermi level in the leads. The conductance will increase whenever the
energy level in the dot is aligned with the Fermi levels of the leads, and decrease in between, forming the so called Coulomb oscillations. Resistance oscillation measurements where performed in an $^3$He cryostat at 300\,mK using a low frequency lock-in
technique with 200\,$\mu V$ ac voltage applied between source (S) and drain (D) contacts. Figure \ref{Coul_osz} shows typical resistance oscillations observed by sweeping the gate voltages of gates G1 and G2 and keeping the voltages of the gates G3 and G4 fixed. Resistance minima are clearly separated by regions of increased resistance.

\begin{figure}
\begin{center}\leavevmode
\includegraphics[width=0.95\linewidth]{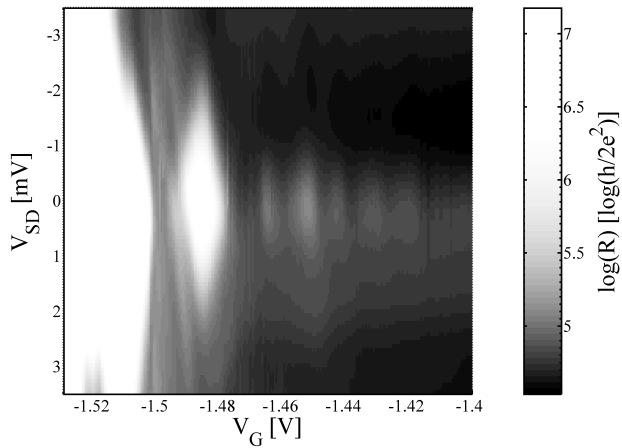}
\caption{Quantum dot spectrum taken at 300mK. }
\label{Diamond}
\end{center}\end{figure}

By measuring the conductance as a function of the plunger gate
voltage and an additional dc voltage applied between source and
drain contacts, one can obtain the quantum dot spectrum, resulting
in Coulomb blockade "diamonds", as shown in figure \ref{Diamond}.
Well-behaved Coulomb blockade diamonds were measured up to an temperature of 1\,K, which
was the maximum reachable temperature in the measurement
apparatus.

\begin{figure}
\begin{center}\leavevmode
\includegraphics[width=0.7\linewidth]{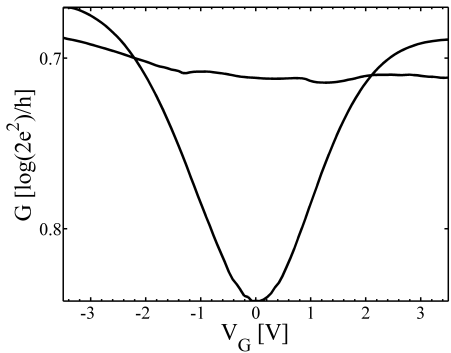}
\caption{Conductivity as a function of the Source-Drain voltage for two combinations of fixed gate voltages. The lower curve shows
    the dot in an nonconducting state. In the upper curve an energy level is aligned with the Fermi sea at no bias.}    
\label{SDcurve}
\end{center}\end{figure}

From the size and shape of the diamonds electrical properties such as capacitances of the gates and leads with respect to the dot can be analyzed. One can estimate the dot capacities and thus the size of the dot by analyzing the distance of neighboring Coulomb diamonds and their slopes. This analysis yields a gate capacity of 5.6\,aF and a source capacity of 34\,aF. The total dot capacity of 56\,aF gives an estimated dot radius $R$ of about 650\,\AA{}, when assuming that the dot capacity is the same as a metallic disc with a capacity of  $C=8\epsilon\epsilon{}_0R$. $\epsilon{}_0$ is the permittivity of vacuum and $\epsilon{}$ the relative permittivity of silicon. Therefor the electron number in the dot can be estimated to be about 40 by using the measured electron density of $3.2\times{}10^{11}\,cm^{-2}$.\\

In summary, a lateral quantum dot has been fabricated on modulation doped Si/SiGe heterostructures with a split-gate technique. The quantum dot spectrum has been measured up to an temperature of 1\,K. These results show that SET functionality can be achieved in modulation-doped Si/SiGe heterostructures with a standard split-gate approach that can easily be integrated into an array of coupled SETs for spintronic applications, as suggested in Ref. \cite{friesen}.\\

The financial support of FWF and GMe, is gratefully acknowledged (both Vienna, Austria).\\ \\
Note: Paper was presented at EP2DS-16 July 12, 2005

\end{document}